\documentclass[twocolumn,preprintnumbers,amsmath,amssymb]{revtex4-2}

\usepackage{graphicx}  
\usepackage{sidecap}
\begin{document}
\title{
 Effect of Coulomb carrier drag 
 and terahertz plasma instability in p$^+$-p-i-n-n$^+$ graphene tunneling transistor structures  
}
\author{V. Ryzhii$^{1,2,a)}$,   M. Ryzhii$^3$, A. Satou$^1$, T.~Otsuji$^1$, V. Mitin$^4$, and M. S. Shur$^5$}
\address{
$^1$Research Institute of Electrical Communication, Tohoku University, Sendai 980-8577, Japan\\
$^2$Institute of Ultra High Frequency Semiconductor Electronics of RAS,
 Moscow 117105, Russia\\
 $^3$Department of Computer Science and Engineering, University of Aizu, Aizu-Wakamatsu 965-8580, Japan\\
$^4$ Department of Electrical Engineering,~University at Buffalo,~SUNY, Buffalo,~ New York 14260 USA\\
$^5$ Department of Electrical,~Computer,~and~Systems~Engineering, Rensselaer Polytechnic Institute,~Troy,~New York 12180, USA\\
$^{a)}$Author to whom correspondence should be addressed: v-ryzhii@riec.tohoku.ac.jp
}
 \begin{abstract} 
We evaluate the influence of the Coulomb   drag of the electrons and holes in
the gated  n- and p-regions by the ballistic electrons and holes  generated  in the depleted i-region due to the interband tunneling
on the current-voltage characteristics and impedance of the p$^+$-p-i-n-n$^+$ graphene tunneling  transistor structures  (GTTSs). 
The  drag leads to a current  amplification in the 
gated  n- and p-regions and  a positive feedback between the   amplified dragged  current and the injected tunneling current.
A sufficiently strong  drag can result in 
 the negative real part of the GTTS impedance enabling the plasma instability and the self-excitation of the plasma oscillations in the terahertz (THz) frequency range. This effect might be used for the generation of the THz radiation.    
\end{abstract} 
\maketitle
%


\begin{figure}[t]
\centering
\includegraphics[width=7.0cm]{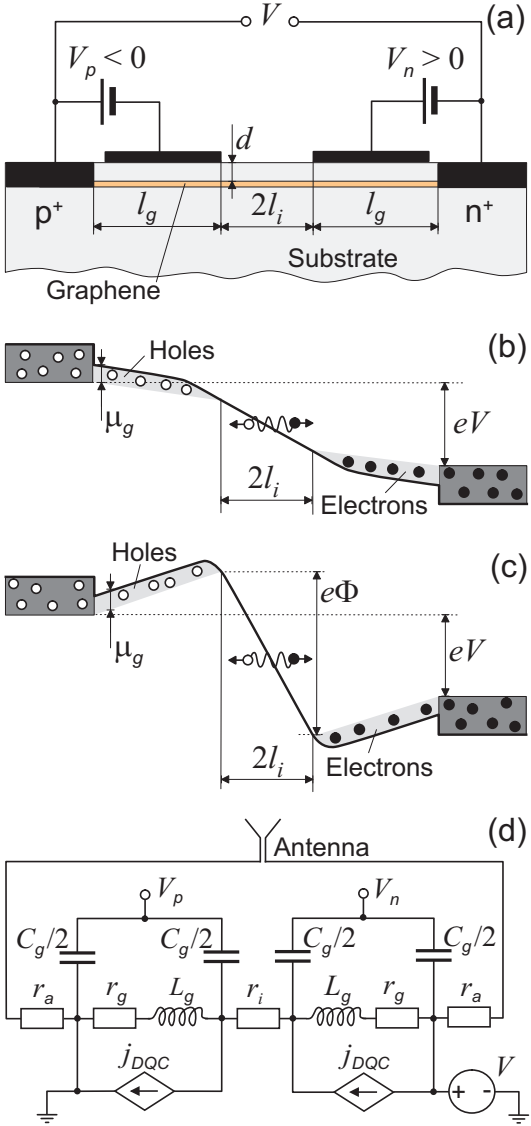}
\caption{ (a) Cross-section,  band diagrams  of a lateral  p$^+$-i-n$^+$  GTTS 
with ballistic  transport of the tunneling generated electrons and holes in i-region (b) in the absence of drag current, $\Phi < V$,  and (c) at a strong Coulomb carrier drag, $\Phi > V$.  
(d)  GTTS  equivalent circuit including an antenna with radiation resistance $2r_l$ ($r_i$ is the i-region resistance, $R_g$, $L_g$, and $C_g$ are the gated region resistance, inductance and capacitance, respectively). The current source describes the DQC current.
}
\label{F1}
\end{figure}

\section{Introduction}

The interband Zener-Klein tunneling 
in the depleted i-region in the p-i-n graphene diodes  leads to the generation of the holes and electrons  with holes propagating primarily
in the direction of the built-in electric field and and electrons propagating in the opposite direction~\cite{1,2,3,4}. If 
 the i-region sufficiently short,
the scattering of the carriers generated in this region on the impurities and acoustic phonons  is ineffective. As a result, 
the generated carriers propagate across such a region ballistically with the directed velocity close to the characteristic velocity $v_W \simeq 10^8$~cm/s.
This very fast carrier transit can be used in different ultra-high frequency
devices operating at the room temperature. 
In particular, the p-i-n graphene diodes and the p$^+$-p-i-n$^+$ 
graphene tunneling transistor structures (GTTSs) can exhibit the negative dynamic conductivity~\cite{5,6},
associated with the transit-time effect at the frequencies close to the inverse transit time $v_W/2l_i$ (where $2l_i$ is the depleted i-region length), which is in the range of several terahertz (THz) for submicron dimensions.
Recently~\cite{7,8}, we demonstrated that in graphene-based  structures the ballistic carrier injected into
the gated n-region (or p-region) can effectively drag~\cite{9,10,11} the equilibrium carriers. Due to the specifics of the Coulomb carrier-carrier scattering in G-layers having the two-dimensional linear dispersion~\cite{9,10,11,12,13,14}, the drag can be accompanied by a marked 
current amplification. This is because the injected ballistic carriers (BCs) colliding with the equilibrium carriers (ECs) transfer to the latter their momenta while keeping their
 directed velocity. 
 As demonstrated previously~\cite{8}, the Coulomb drag effect in the G-based n$^+$-i-n-n$^+$ field-effect transistors  with the injection of ballistic electrons (GBFETs) from the  source can enable the current-driven plasma instability and the self-excitation of the source-drain current, leading to the emission of the THz radiation.

In this paper we consider the reversed biased p$^+$-p-i-n-n$^+$ GTTSs with the gated p- and n-regions induced by the negative and positive gate voltages $V_p < 0$ and $V_n > 0$, respectively, i.e., by
the so-called, electrostatic doping.

Figure 1~(a) shows the GTTSs at the applied gate  voltages, $V_p$ and $V_n$ ($-V_p =V_n = V_g >0$)
and the reverse bias voltage, $V$ between the side  contacts.
The voltage results in the depletion of the i-region and creation of the sufficiently strong lateral electric field in this region enabling the effective
interband tunneling.
Figures~1(b) and 1(c) show qualitatively the GTTS band diagrams in the case of negligible Coulomb  carrier drag (as in ~\cite{5}) and the case when such a drag is substantial.
In the first case, the potential distribution across the  p- and n-regions is nearly flat due to relatively high conductivity of these regions.
However, since the drag of the equilibrium carrier by the injected ballistic carriers 
pushes  out fractions of the equilibrium carriers to the side contacts, to compensate
these carriers displaced from the gated regions (or prevent such a displacement)
lateral build-in electric fields  along the p- and n- regions arise (corresponding to a marked potential inclination in the gated region).
The GTTSs under consideration differ from the forward-biased GBFETs studied previously by different mechanism of ballistic carriers generation (the interband Zener-Klein tunneling in the GTTSs versus the thermionic space-charge limited injection in the GBFETs) and, hence, different energy distributions of the carrier injected into the gated regions. This leads to different device characteristics, in particular, to the different criteria 
of the plasma instability. 

Using the developed device model, we calculate the GTTS dc current-voltage ({\it I-V}) characteristics and the small-signal impedance. In particular, we demonstrate
that the real part of the GTTS impedance can be negative in a certain range of the THz frequencies  and  the device structural parameters and the bias voltages. The impedance imaginary part can turn into zero at the plasma frequency. These conditions can correspond to the plasma instability and the self-excitation
of plasma oscillations converted to the emitted THz radiation.

\section{Device model}

Considering the GTTSs, we use the GTTS equivalent circuit shown in Fig.~1(d).
Similar G-based structures with the enhanced carrier mobility were fabricated, experimentally studied, and well documented in the literature (see, for example,~\cite{15,16,17,18,19,20,21,22,23}).

 In contrast to the  monopolar  n$^+$-i-n-n$^+$ GBFETs using the drag effect associated with the ballistic electrons thermally injected from
the n$^+$ contact (the space-charge limited injection) 
into the i-region and then into the      
gated n-region~\cite{7,8}, here we consider the GTTSs with
the interband  tunneling electron and hole generation  across the entire i-region (approximately spatially uniform).
We analyze the drag by the ballistic carriers with the energies distributed in a wide range (from approximately zero energy  for the electrons/holes generated
near the n-region/p-region to the energy $\varepsilon = eV$, where $e =|e|$ is the electron charge). We account for 
the asymmetric potential distribution and the contribution of the carriers of both types to the device characteristics.

We assume that:\\
(i) The length, $2l_i$, of the  depleted i-region is  sufficiently short allowing for the ballistic motion of the injected electrons: 
$2l_i \ll \tau_{i}/v_W$, where $\tau_i$ is the characteristic time of the BC scattering on the disorder (acoustic phonons and impurities). 
 As demonstrated~\cite{24}, the latter condition at $2l_i \sim 1~\mu$m 
 can be satisfied at room  and  lower temperatures
 in the G-layers encapsulated in hBN~\cite{25};\\
(ii) The characteristic time, $\tau_{cc}$,  of the ballistic carriers
(BCs) scattering  on the equilibrium carriers (ECs) is much shorter than the EC scattering time, $\tau_g$, in the gated p- and n-regions: $\tau_{cc} \ll \tau_{g}$ (with 
the characteristic times of the BC and EC scattering on the disorder being approximately equal to each other, i.e., $\tau_g \sim \tau_i$). In this case, a substantial fraction of  the BCs momentum can be transferred to the ECs converting them into the dragged ECs (DECs);\\
(iii) The BCs acquiring the energy exceeding the 
threshold of the optical phonon emission  (about its energy $\hbar\omega_0 \simeq 0.2$~eV)
are scattered 
primarily in the gated regions with a relatively small characteristic time $\tau_{0}$. A fraction of the optical phonons emitted by the BCs in the depletion region is small if the potential drop across this region is smaller than $ \hbar\omega_0$;\\
(iv) Due to a short length of the i-region, the delay of the carrier transit, the capacitance, and the kinetic inductance of this region
are  disregarded.\\

At the carrier densities $\Sigma_g \simeq 1\times(10^{12} - 10^{13})$~cm$^{-2}$ and  temperature $T \lesssim 300$~K for the  energy, $\varepsilon_{BC}$,  of the BCs injected into the p- and n-regions one can assume  $\tau_{cc} \lesssim 0.1$~ps.
 The characteristic times of the scattering on disorder and on optical phonons
 are estimated as $\tau_g \simeq (1 - 2)$~ps and  $\tau_{0} \simeq (0.5 - 1.0)$~ps ~\cite{26}. 
 The use of the electrostatic doping of the p- and n-regions allows to
minimize the ratio $\tau_{cc}/\tau_{g}$. The characteristic time of the optical phonon spontaneous emission is set to be $\tau_{0} \simeq (0.5 - 1.0)$~ps.

As seen from the GTTS  equivalent circuit shown in Fig.~1(d), one  needs
to equalize  the BC current across the i-region (equal to the terminal current) and the net currents across
 the p- and n-regions (the Kirchhoff circuit law). As a result, we  arrive at the following equation:

\begin{equation}\label{eq1}
J_{BC}  = J_{DEC} + J_{EC} + J_{DP}.
 \end{equation}
Here  $J_{BC}$, $J_{EC}$,  $J_{DP}$, and $J_{DEC}$  are the  densities of the BC current injected into the gated n-region (or the gated p-region), the QC current in each gated region, the displacement current, and the current of the DeCs, respectively.  
These quantities are given by the following equations:

\begin{equation}\label{eq2}
J_{BC} = \ae_i\,\Phi^{3/2},
\end{equation}

\begin{equation}\label{eq3}
J_{EC} = \sigma_g\frac{(V-\Phi)}{2l_g} - \frac{{\mathcal L}_g}{l_g}
\frac{d\,J_{QC}}{dt}, 
\end{equation}
\begin{equation}\label{eq4}
J_{DP} = c_g\,
\frac{d(V - \Phi)}{dt}.
\end{equation} 
Here $\ae_i = \displaystyle\frac{e^{5/2}}{2\pi^2\hbar^{3/2}\sqrt{2l_iv_W}}$~(see~\cite{1,2,3,4,5,6}), 
$\sigma_g = \displaystyle\frac{e^2\Sigma_g \tau_g}{m}$ is the drift (Drude) conductivity, 
$c_g = \displaystyle\frac{l_g\kappa}{8\pi\,d}$, 
 and ${\mathcal L}_g$ are  the capacitance and the kinetic carrier inductance (per unit withs of the GTTS in lateral direction perpendicular to the terminal current)) of the gated-region, 
  $d$ and $\kappa$ are   gate layer thickness and its dielectric constant,
and  $m_g = \mu_g/v_W^2 \simeq \hbar\sqrt{\pi\Sigma_g}/v_W^2$ and  $\mu_g \simeq  \hbar\,v_W\sqrt{\pi\Sigma_g}$ 
are the QC fictitious effective mass and the Fermi energy.
The quantities
$\Phi$ and $(V - \Phi)/2$ are the potential drops across the i- and each gated-regions.


\section{DEC current}

Calculating the DEC current, we use the approach similar to that
in~\cite{7}, accounting for the features of 
 the BC injection into the gated regions in the GTTSs. 
As a result, we arrive at the following equation: 

\begin{eqnarray}\label{eq5}
J_{DEC} \simeq  b\frac{J_{BC}^{5/3}}{J_0^{2/3}}e^{-K}.
\end{eqnarray}
Here 

\begin{equation}\label{eq6}
K =\frac{K_0}{2}\frac{[(e\Phi - \hbar\omega_0 + \mu_g)^2 - \mu_g^2)]}{\hbar^2\omega_0^2}\Theta(e\Phi - \hbar\omega_0),
\end{equation}
$K_g = l_g/v_W\tau_g$, $K_0 =l_g/v_W\tau_{0}$, 
$K_{cc} = l_g/v_W\tau_{cc}$, $K_g = l_g/v_W\tau_g$ with
$K_g < K_0 \ll K_{cc}$,  
The factor of two in the denominator is because the carriers injected into the gated region have the average energy equal to $e\Phi/2$.
The dependence of $K$ on the Fermi energy $\mu_g$ describes
the effect of the gated region population by electrons and holes on the energy threshold of the optical phonon emission, which is reflected by the unity step function $\Theta$ with threshold energy  $\hbar\omega_0 + \mu_g$. 
Equation~(6) accounts for this effect and the linear energy dependence of the density of states. 
The Coulomb carrier drag factor $b$ is given by

\begin{eqnarray}\label{eq7}
b = \frac{\hbar\omega_0}{3T}\frac{{\cal F}_1(\mu_g/T)}{{\cal F}_2(\mu_g/T)}
e^{-K_{g}}
 \simeq \frac{\hbar\omega_0}{2\mu_g}
e^{-K_{g}} \lesssim \frac{\hbar\omega_0}{2\mu_g}.
\end{eqnarray}
with $T$ being the temperature (in energy units).
 In Eq.~(5), we have introduced the characteristic current density
\begin{eqnarray}\label{eq8}
J_0 = \ae_i \biggl(\frac{\hbar\omega_0}{e}\biggr)^{3/2}.
\end{eqnarray}

The quantity $J_{DQE}$ is different from that calculated previously~\cite{7}
for the case of virtually monoenergetic BCs.

\section{DC characteristics}
 
At the dc bias voltage $V = {\overline V}$ applied between the side contacts, $\Phi = {\overline \Phi}=const$,
and the density of the terminal current
$J_{BC} = {\overline J_{BC}} =const$. 
Introducing the normalized current density ${\overline J = {\overline J_{BC}}/J_0}$, 
from Eq.~(1) we arrive at the following  equation relating ${\overline J}$ and ${\overline V}$ (i.e., the GTTS {\it I-V} characteristic):

\begin{eqnarray}\label{eq9}
 {\overline J} + \eta{\overline  J}^{2/3} -b {\overline J}^{5/3}
 e^{-K({\overline J})}
 =\eta \frac{{\overline V}}{V_0},
\end{eqnarray}
where 

\begin{eqnarray}\label{eq10}
\eta =\frac{\sigma_g}{2l_g\ae_i \sqrt{V_0}} = \frac{\pi\tau_g\mu_g}{\hbar}\sqrt{\frac{2l_iv_W}{l_g^2\omega_0}}
\end{eqnarray}
is the ratio of the i- and g-regions resistances  and $V_0 = \hbar\omega_0/e$.
In Eq.~(9), we have expressed the dc potential drop across the depleted region
${\overline \Phi}$ via ${\overline J}$: ${\overline \Phi} = {\overline J}V_0$.
Taking this and Eq.~(6) into account, we obtain

\begin{eqnarray}\label{eq11}
K({\overline J}) = \frac{K_0}{2}[({\overline J}^{2/3} - 1 +F )^2 - F^2]\cdot\Theta({\overline J} - 1).
\end{eqnarray} 

Analyzing Eq.~(9), one can find that
at $b\Lambda \leq 1 +2\eta/3$, the GTTS {\it I-Vs} are monotonic.
In this case, ${\overline \Phi} < {\overline V}$ [see Fig.~1(b)].
In the opposite case when $b\Lambda > 1 +2\eta/3$, the GTTS {\it I-Vs} become of the S-shape analogously to those demonstrated previously~\cite{7}
(in the devices with different type of the BC injection). In the latter case, ${\overline \Phi} > {\overline V}$ [see Fig.~1(c)].
Here 

\begin{eqnarray}\label{eq12}
\Lambda = \frac{d[J^{5/2}e^{- K(J)}]}{dJ}
\biggr|_{J ={\overline J}}.
\end{eqnarray}
When ${\overline J} \lesssim 1$,
as follows from Eq. (11),  $K({\overline J}) \simeq const$, and $\Lambda \simeq 5{\overline J}^{3/2}/2$.
Considering this, the formation of the S-shaped {\it I-V} characteristics requires

\begin{eqnarray}\label{eq13}
\frac{5}{2}b > 1 + \frac{2}{3}\eta,
\end{eqnarray}
while the sufficient condition of the {\it I-V} monotonic shape
(despite the drag effect) is presented as

\begin{eqnarray}\label{eq14}
 \frac{5}{2}b < 1 + \frac{2}{3}\eta.
\end{eqnarray}

Figure~2 shows the {\it I-V} characteristics calculated using Eqs.~(9)
and (11) 
for different 
structura1 parameters. We  set  $\tau_0 = 0.5$~ps, $2l_i = 0.2~\mu$m, $l_g = 0.5~\mu$m, $\tau_g = 1$~ps for  $\mu_g =  50$~meV ($K_g = 0.5$) and $\tau_g = 2$~ps for $\mu_g = 15$~meV ($K_g = 0.25$). This implies that $K_{0} = 2.0$. 
The {\it I-V} characteristics for  $\mu_g =  50$~meV, which can be
 attributed to $T = 300$~K ($\mu_g/T \gtrsim 2$),   is monotonically rising. Very similar characteristics (virtually 
undistinguishable)
are obtained for the  Fermi energy range $\mu_g = 40 - 60$~meV 
with  $b \simeq 1.0 - 1.5$ and $\eta \simeq 10 - 15$. The latter implies  that inequality (14) is well satisfied. Therefore  the {\it I-V} characteristics for $\mu_g = 40 - 60$~meV  
are monotonic. 
For comparison, we calculated  also the  {\it I-V} characteristics corresponding to 
 lowered temperatures setting  the Fermi energy $\mu_g = 15$~meV  (the Fermi energy can be decreased by decreasing of the gate voltage $V_g$).
These data can be conditionally ascribed to $T = 77$~K with $\mu_g/T \gtrsim 2$.
An example of the pertinent {\it I-V} characteristics  also shown in Fig.~2 exhibits the pronounced S-shape.  This characteristics corresponds to the parameters $b$ and $\eta$
 satisfying an inequality~(13), which is  opposite to inequality~(14), i.e., relatively large $b$ and small $\eta$;
in this case $b\simeq 5.19$ and $\eta \simeq 7.54$.

Figure 3 shows the quantity $b\Lambda$ as a function of the normalized bias current calculated using Eqs.~(11) and (12).
As seen from Fig.~3, the value of  $b\Lambda$ explicitly  depends on the structural parameters
and the bias current. It is crucial that  $b\Lambda$ could markedly exceed unity particularly at $\overline J \simeq 1$ for the parameters chosen above. Small fissures in the $b\Lambda$ versus ${\overline J}$ plots near ${\overline J} = 1$ are attributed to an inclusion of the optical phonon emission at this point.
The optical phonon emission, which starts from ${\overline J} \simeq 1$ (and ${\overline \Phi} \simeq 1$), results in a drop of $b\Lambda$
at larger values of ${\overline J}$. This is because an increase in  ${\overline J}$
leads to relatively smooth increase of $K(\overline J)$ (and ${\overline \Phi}$) beyond the optical phonon emission threshold [see Eq.~(11)].
When ${\overline J} \gtrsim 1$ (${\overline \Phi} \gtrsim 1$), only the BCs generated near the gated regions have sufficient energy to emit an optical phonon. The BCs generated in the bulk of the i-region are able to do that only
at  higher values of ${\overline J}$. This is in contrast with the situation in the GBFETs~\cite{7,8}, where
the optical phonon emission threshold is rather sharp. 

One needs to mention that a difference in the values of $\Lambda$ at different Fermi energies
under consideration is rather small, while the quantity $b\Lambda$ is fairly sensitive to the Fermi energy. This is because of $b \propto \mu_g^{-1}$.

\begin{figure}[t]
\centering
\includegraphics[width=8.0cm]{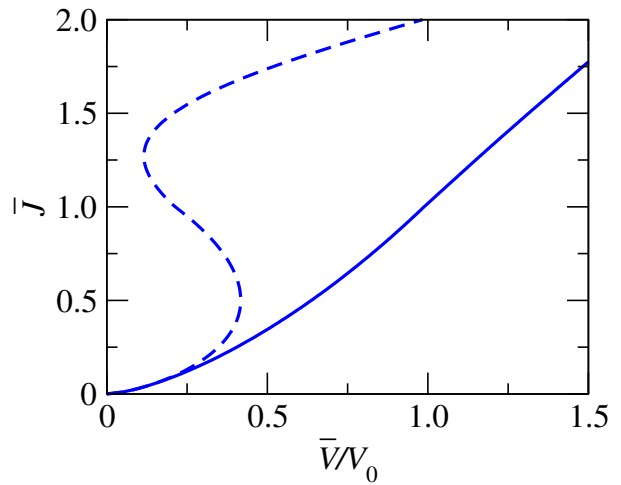}
\caption{ GTTS {\it I-V} characteristics  for different 
structura1 parameters:  $l_g = 0.5~\mu$m, $\mu_g =  50$~meV, and $\tau_g = 1$~ps - solid line (monotonic) and $l_g = 0.5~\mu$m, $\mu_g =  15$~meV, and $\tau_g = 2$~ps - dashed line (S-shaped).
} 
\label{F2}
\end{figure}

\begin{figure}[t]
\centering

\includegraphics[width=8.0cm]{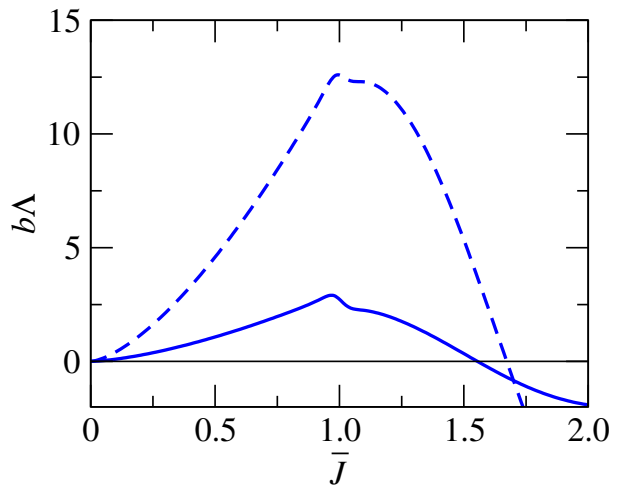}
\caption{Parameter $b\Lambda$ versus  normalized bias current ${\overline J}$ for different 
structura1 parameters: $l_g = 0.5~\mu$m, $\mu_g =  50$~meV, and $\tau_g = 1$~ps - solid line and $l_g = 0.5~\mu$m, $\mu_g =  15$~meV, and $\tau_g = 2$~ps - dashed line. 
} 
\label{F3}
\end{figure}

\section{AC current and impedance}
 
 When $V = {\overline V} + \delta V\exp(-i\omega t)$, where $\delta V$ and $\omega$ are the signal amplitude and frequency,  
the potential drop across the depleted region and the normalized 
 terminal current comprises the ac component $\delta \Phi$ and  $\delta J = \delta J_{BC}/J_0$, respectively. Using Eqs.~(3) - (5), the ac normalized current components  $\delta J_{DQC}$
 can be presented as

\begin{eqnarray}\label{eq15}
\frac{\delta J_{QC} + \delta J_{DP}}{J_0}\simeq 
\frac{2\eta}{3{\overline J}^{1/3}}\biggl[- \frac{1}{(1-i\omega\tau_g)}
+i\omega\tau_{rc}\biggr]\delta J
\nonumber\\
+\eta \biggl[-\frac{1}{(1-i\omega\tau_g)} +i\omega \tau_{rc}\biggr]\frac{\delta V}{V_0}
\end{eqnarray}
with 
$\tau_{rc} = (2l_gc_g/\sigma_g)$ being the gate recharging time, where the relation ${\mathcal L}_g
=  (l_gm_g/e^2\Sigma_g) = l_g\tau_g$ is used.

Using Eq.~(5), the ac normalized drag current can be presented as

\begin{eqnarray}\label{eq16}
\delta J_{DQC} \simeq b \Lambda \delta J.
\end{eqnarray}
Considering this, the linearized version of Eq.~(1) with Eqs.~(15) and (16) yield

\begin{eqnarray}\label{eq17}
\delta J \biggl\{ 1 - b\Lambda + \frac{2\eta}{3{\overline J}^{1/3}}\biggl[ \frac{1}{(1-i\omega\tau_g)}
-i\omega\tau_{rc}\biggr]
\biggr\}\nonumber\\
 =  \eta \biggl[\frac{1}{(1-
 i\omega\tau_g)} -i\omega \tau_{rc}\biggr]\frac{\delta V}{V_0}.
\end{eqnarray}
Equation~(17) can be presented in the following form:

\begin{eqnarray}\label{eq18}
\delta J =\frac{1+i{\mathcal F}_{\omega}}
{\displaystyle\biggl[\frac{3{\overline J}^{1/3}(1 -b\Lambda)}{2\eta}(1+\omega^2\tau_g^2) +1+i{\mathcal F}_{\omega}\biggr]
}
 \frac{\delta V}{V_0}.
\end{eqnarray}
Here

\begin{eqnarray}\label{eq19}
{\mathcal F}_{\omega} = \frac{\omega}{\Omega^2\tau_g}[(\Omega^2 - \omega^2)\tau_g^2 - 1],
\end{eqnarray}
where

\begin{eqnarray}\label{eq20}
\Omega = \frac{1}{\sqrt{\tau_g\tau_{rc}}} = \sqrt{\frac{8\pi\,e^2\Sigma_gd}{\kappa\,m_g\,l_g^2}} = \frac{e}{\hbar\l_g}\sqrt{\frac{8\mu_gd}{\kappa}}
\end{eqnarray}
is the plasma frequency of the gated carrier system.
One can see that $\Omega = 1/\sqrt{c_g{\mathcal L}_g}$.

Taking into account the load resistances, we arrive at the following expression for the GTTS impedance $Z_{\omega}$ normalized by the depleted i-region resistance $r_i$:

\begin{eqnarray}\label{eq21}
\frac{Z_{\omega}}{r_i} =
\displaystyle\frac{3{\overline J}^{1/3}(1 -b\Lambda)(1+\omega^2\tau_g^2)}
{2\eta(1+i{\mathcal F}_{\omega})} + 1 +\rho,
\end{eqnarray}
where 
\begin{eqnarray}\label{eq22}
r_iH = \frac{1}{\ae_i\sqrt{V_0}} = 2\pi^2 \biggl(\frac{\hbar\,v_W}{e^2}\biggr)\sqrt{\frac{2l_i}{\omega_0\,v_W}},
\end{eqnarray}
 $\rho = 2r_{a}/r_i$, and $2r_{a}$ is  the antenna  radiation resistance)
 [see Fig.~1(d)], and $H$ is the GTTS width.
 In the range ${\overline J} \leq 1$, Eqs.~(13) and (21) result in
 
\begin{figure}[t]
\centering
\includegraphics[width=8.0cm]{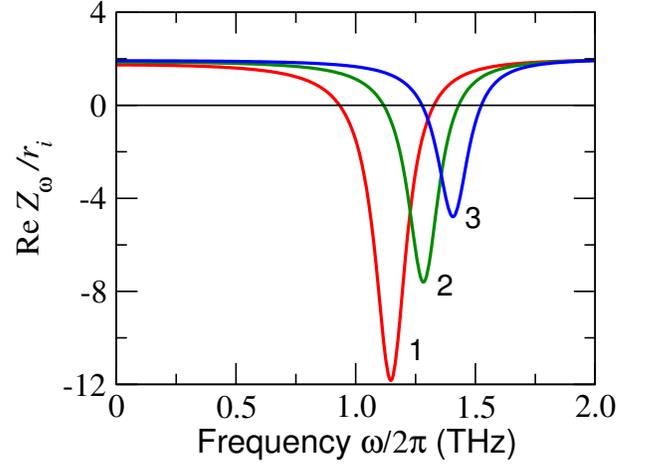}
\caption{The   real part Re~$Z_{\omega}/r_i$  of the GTTS impedance
versus  signal frequency $f =\omega/2\pi$ at $\tau_g = 1$~ps, $l_g = 0.5~\mu$m,
 and $\rho = 1$, and ${\overline J} =1$: 
for different structural parameters:
   1 - $\mu_g = 40$~meV and $\Omega/2\pi$ = 1.15 THz),
 2 -  $\mu_g = 50$~meV and $\Omega/2\pi = 1.3$~THz, and 
 3 - $\mu_g = 60$~meV and $\Omega/2\pi = 1.4$~THz.
}
\label{F4}
\end{figure}
\begin{figure}[t]
\centering
\includegraphics[width=8.0cm]{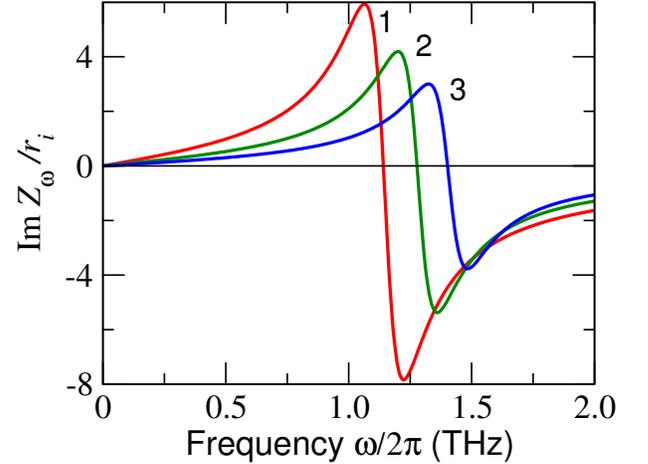}
\caption{The imaginary part Im~$Z_{\omega}/r_i$  of the GTTS impedance
versus  signal frequency $f =\omega/2\pi$ (parameters are the same as for Fig.~4).
}
\label{F5}
\end{figure}

\begin{eqnarray}\label{eq23}
\frac{Z_{\omega}}{r_i} =
\displaystyle\frac{3 {\overline J}^{1/3}}{2\eta}\biggl(1 - \frac{5b}{2}{\overline J}^{2/3}\biggr)\frac{(1+\omega^2\tau_g^2)}
{(1+i{\mathcal F}_{\omega})} + 1 +\rho.
\end{eqnarray}
As follows from Eqs.~(21) and (23), the real part of the GTTS impedance can become negative 
if

\begin{eqnarray}\label{eq24}
\frac{5}{2}b > 1 + \frac{2}{3}\frac{\eta}{(1 +\omega^2\tau_g^2)}.
\end{eqnarray}
For high signal frequencies ($\omega^2\tau_g^2 \gg 1$), the condition
described by inequality (24) is much more liberal that the condition~(13) of the S-shaped {\it I-V} characteristics. This implies that the real part of the GTTS impedance can be negative at the monotonic dc {\it I-V} characteristics. It is remarkable that at moderate values of $b$ satisfying inequality~(24),  but
insufficient to satisfy condition (13), $|\delta \Phi|$ can exceed $\delta V$, while
${\overline \Phi} < {\overline V}$. This corresponds to the potential profile shown in Fig.~1(c) for the signal component but to the potential profile shown in Fig.~1(b) for the dc component.
The inclusion of the load resistance somewhat changes the above criteria, but
not too much (see below).

Condition~(24) for $(1 +\omega^2\tau_g^2) \gg \eta$ yields $b > 2.5$, when the drag effect is essential but not particularly  strong. Assuming $\omega/2\pi = 1$~THz and $\tau_g = 1$~ps, we obtain $(1+\omega^2\tau_g^2) = 1+4\pi^2 \simeq 40.5$.

For  $2l_i = 0.1~\mu$m, Eq.~(8) yields $J_0 \simeq 1.5$~A/cm. Setting the GTTS width equal to
 $H = (5 - 10) ~\mu$m, we find $r_i/H \simeq (133 - 266)~\Omega$, with  $|{\rm Re}~Z_{\omega}|$ markedly exceeding the latter values.

Figures 4 and 5 show the frequency dependences of the normalized  real and imaginary parts of the GTTS impedance calculated using Eq.~(21) with Eq.~(12)  for different structural parameters.

\section{Plasma resonance and plasma instability}

At the signal frequency $\omega =\sqrt{(\Omega^2 - \tau_g^{-2})} = \Omega_p$ corresponding to the plasmonic resonance, ${\mathcal F}|_{\omega = \Omega_p} = 0$.
In this case, Eq.~(21) yields for the resonant impedance $Z_{\omega}|_{\omega = \Omega_p} =Z_{\Omega}$

\begin{eqnarray}\label{eq25}
\frac{Z_{\Omega}}{r_i} =
\displaystyle\frac{3{\overline J}^{1/3}(1 - b\Lambda)\Omega^2\tau_g^2}
{2\eta} + 1 +\rho.
\end{eqnarray}

Figure~6 shows the real part of the GTTS impedance at the signal frequency coinciding with the plasma frequency $\Omega$ (the imaginary part is equal to zero) as a function of the normalized bias current ${\overline J}$. 
One can see  $|Z_{\Omega}| $ can be fairly large, which is beneficial for the plasma instability leading to an effective THz emission (see below). Moreover, the range of the bias currents corresponding to Re~$Z_{\Omega} < 0$
markedly extends into the range  ${\overline J} > 1$. This is attributed to a smooth roll-off of parameter $\Lambda$ in this range as
seen in Fig.~3.

\begin{figure}[t]
\centering
\includegraphics[width=8.0cm]{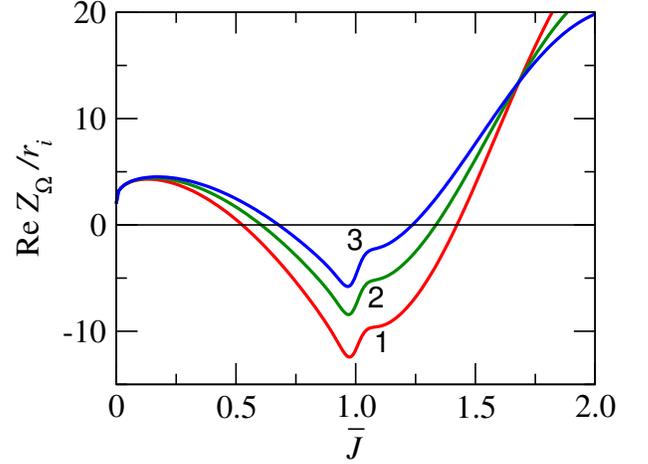}
\caption{Real part of the normalized GTTS impedance  Re~$Z_{\Omega}/r_i$ 
at the plasma frequency ($\omega = \Omega$)
versus normalized bias current ${\overline J}$ for different plasma frequencies:  1 -  $\Omega/2\pi$ = 1.15 THz),
 2 -$\Omega/2\pi = 1.3$~THz, and 
 3 - $\Omega/2\pi = 1.4$~THz. Other parameters are the same as in Figs.~4 and 5.
}
\label{F6}
\end{figure}

As known (see, for example,~\cite{27}), the current in the system with the negative
real part of impedance and the imaginary part of the latter changing sign can be unstable 
resulting in the self-excitation of the carrier density and current high-frequency oscillations
(the plasma instability~\cite{28}). Such  current oscillations feeding an antenna can lead to the radiation emission. According to Figs.~4 and 5, the frequency of the emitted radiation  
can be in the THz range.

The conditions of the plasma instability and the growth rate of the plasma and current oscillations can be found from the dispersion equation $Z_{\omega} = 0$ with the complex frequency $\omega = \omega^{\prime} + i \omega^{\prime\prime}$, where $\omega^{\prime\prime}$ 
represents the growth rate. Using Eq.~(23), at $\Omega^2\tau_g^2 \gg 1$ and $\Omega^2\tau_g^2 > \eta$
(so that $\Omega_p \simeq \Omega$)
from the equations Re $Z_{\omega^{\prime}+ i\omega^{\prime\prime}} = 0$ and 
Im $Z_{\omega^{\prime}+ i\omega^{\prime\prime}} = 0$, we arrive at

\begin{eqnarray}\label{eq26}
\frac{\omega^{\prime}}{\Omega} \simeq 
1 - \frac{5b}{4\eta(1+\rho)} ({\overline J} - {\overline J}_{th})\nonumber\\
\simeq 1 - \frac{5b}{4\eta(1+\rho)^2} \frac{({\overline V} - {\overline V}_{th})}{V_0},
\end{eqnarray}

\begin{eqnarray}\label{eq27}
\frac{\omega^{\prime\prime}}{\Omega} \simeq 
\frac{5b\Omega\tau_g}{4\eta(1+\rho)}({\overline J} - {\overline J}_{th})
\nonumber\\
\simeq \frac{5b\Omega\tau_g}{4\eta(1+\rho)^2}\frac{({\overline V} - 
{\overline V}_{th})}{V_0}.
\end{eqnarray}  
Here $V_{th}/V_0 = (1+\rho)r_iJ_0{\overline J}_{th}/V_0 = (1+\rho){\overline J}_{th}$, where ${\overline J}_{th}$ represents the normalized threshold current obeying the following equation:

\begin{eqnarray}\label{eq28}
{\overline J}_{th}^{1/3} - \frac{5b}{2} {\overline J}_{th} = -\frac{2\eta(1+\rho)}{3\Omega^2\tau_g^2}.
\end{eqnarray} 
Equation~(28) yields

\begin{eqnarray}\label{eq29}
{\overline J}_{th} \simeq \biggl(\frac{2}{5b}\biggr)^{3/2}\biggl[1 + \biggl(\frac{5b}{2}\biggr)^{1/2}\frac{\eta(1+\rho)}{\Omega^2\tau_g^2}\biggr].
\end{eqnarray} 
Hence  

\begin{eqnarray}\label{eq30}
\frac{{\overline V}_{th}}{V_0} \simeq (1+\rho)\biggl(\frac{2}{5b}\biggr)^{3/2}\biggl[1 + \biggl(\frac{5b}{2}\biggr)^{1/2}\frac{\eta(1+\rho)}{\Omega^2\tau_g^2}\biggr].
\end{eqnarray} 
As follows from Eqs.~(26) and (27), when ${\overline V}$ exceeds the threshold value ${\overline V}_{th}$,
the plasma oscillations become unstable with the growth rate $\omega^{\prime\prime} > 0$.
The plasma instability threshold ($\omega^{\prime\prime} = 0$),
corresponds to $\omega^{\prime} = \Omega_p \simeq \Omega$.

The inequalities $V_{th} < {\overline V} \leq V_0(1+\rho)$, i.e., ${\overline J}_{th} < {\overline J} \leq 1$, and $\omega^{\prime\prime} > 0$ constitute the necessary and sufficient condition of the plasma instability  (compare with inequality~(24), which is the pertinent necessary condition).  
One can see that at a sufficiently large load resistance ($\rho \gg 1$), the plasma instability vanishes,
i.e., the plasma oscillations become damped. 
The proportionality of the plasma oscillations growth to the drag factor $b$ seen in Eq.~(27) shows that the plasma instability under consideration is linked with the drag effect. When $b$ tends to zero, ${\overline J}_{th}$ and ${\overline V}_{th}$
turn to infinity.

Figure~7 illustrates the dependences  $\omega^{\prime}$  and  $\omega^{\prime\prime}$, i.e., the frequency of
the self-excited plasma oscillations and their grows rate versus  $({\overline J} - {\overline J}_{th})$
numerically  calculated (solid lines) using equations Re $Z_{\omega^{\prime}+ i\omega^{\prime\prime}} = 0$ and 
Im $Z_{\omega^{\prime}+ i\omega^{\prime\prime}} = 0$ with $Z_{\omega}$ given by Eq.~(23). The dashed lines in Fig.~7 correspond to an analytic approximation obtained from Eqs.~(26), (27), and (29).

\begin{figure}[t]
\centering
\includegraphics[width=8.0cm]{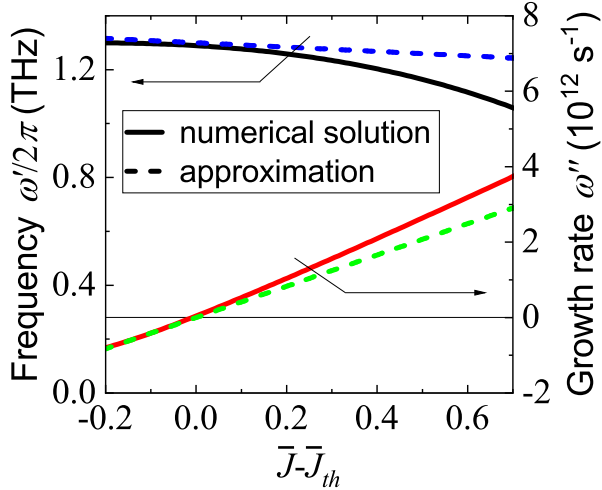}
\caption{Real, $\omega^{\prime}$, and imaginary, $\omega^{\prime\prime}$, parts of the complex  signal frequency
as functions of the deviation of bias current ${\overline J}$ from its instability threshold value 
${\overline J}_{th}$ for  $\tau_g = 1$~ps, $l_g = 0.5~\mu$m, $\mu_g = 50$~meV,  $\Omega/2\pi = 1.3$~THz,
$\rho = 1$ 
(i.e., for $b =1.25$ and $\eta =12.56 $).
}
\label{F7}
\end{figure}

\section{Comments}

Above, as in~\cite{7,8}, we disregarded the delay of the BCs flight across the i-region and, hence, the pertinent transit-time effects~\cite{5,6,29}. Such effects could also be responsible for the oscillation self-excitation  at the transit-time frequency $\omega \sim \omega_{tr} = 2\pi\,v_W/l_i$~\cite{5} and could 
in principle, affect the instability mechanism considered in the present paper.
However, 
at  $\omega < \omega_{tr}$, the omission of the  transit-time delay in the i-region is justified. Indeed,
for the main parameters used in the above calculations $\omega_{tr}/2\pi \simeq 10$~THz, while above $\omega/2\pi, \Omega/2\pi$ were about 1~THz i.e., $\omega, \Omega \ll \omega_{tr}$.

The parameter crucial for the dc {\it I-V} characteristics and high-frequency performance of the devices using the Coulomb carrier drag is the characteristic current density $J_0$.
 Comparing   $J_0$  with the pertinent density for the GFETs $J_0^{(GBFET)}$ ~\cite{7,8} we obtain

$$
\frac{J_0}{J_0^{(GBFET)} }
=
\frac{e^2l_i^{(GBFET)}}{\pi\kappa\hbar\,v_W}\sqrt{\frac{\omega_0}{2l_iv_W}}.
$$
Setting $\kappa = 4 - 6$ and equal lengths of the i-region ($l_i^{GBFET} = 2l_i =0.1~\mu$m) and the gated regions,  we obtain 
$J_0/J_0^{(GBFET)} \simeq 0.7 -1.0$.
This implies that GTTSs and GBFETs exhibit similar characteristics. 

The GTTSs in comparison to the GBFETs are characterized by a smaller drag factor (twice as small
at the same carrier scattering time in the gated region). 
This is because in the former case,
the average energy and momentum of the injected BCs are equal to $eV/2$ and $eV/2v_W$, respectively, i.e., in the GBFETs with the thermionic emission from the source contact the GTTS drag factor is two times smaller (for the same gated region lengths). However, this is compensated by a smoother decrease in $\Lambda$ beyond the optical phonon emission threshold. The latter also leads to a wider span, $\Delta J$, of the bias current in which the real part of the impedance can be negative compared to the GBFETs. 
In particular, the GTTS impedance can be negative at ${\overline J} > 1$, while the GBFET impedance turns to the positive values once ${\overline J} \geq 1$.
The quantity $\Delta J$ determines the characteristic (maximum)
output THz  power emitted by a GTTS. For $2l_i = 0.1~\mu$m, taking into account  $J_0 \simeq 1.5$~A/cm and
assuming the GTTS width $H= 10~\mu$m, $r_i = 200/1.5 \simeq 133~\Omega$, and $\Omega/2\pi = 1.15$~THz. Hence,  as follows from Fig.~6, for $|Z_{\Omega}| = 2r_i \simeq 266~\Omega$ one can find   $H\Delta J \simeq 0.75 HJ_0 \simeq 1.125$~mA. As a result,  for the impedance matched antenna radiation resistance equal to $2r_a = |Z_{\Omega}| = 266~\Omega$,
 the characteristic emitted THz power $P_{\omega} \simeq 337.5~\mu$W, or $P_{\omega}/H \simeq 37.5$~mW/mm.

  The sensitivity of the GTTS characteristics to several parameters opens up  wide opportunities of their optimization.

The fact that in the GTTSs the tunneling injection occurs in  the bulk of the i-region can be considered as an advantage, since it weakens the requirements for the  side contacts quality.

The devices based on the GTTS structures with chemically doped regions can exhibit similar phenomena
to the gated GTTSs with the electrostatic doping of the p- and n-regions.
To avoid excessive scattering of the BCs on the ionized dopants, the acceptors and donors  should not  be placed too close to the channel (selective remote doping). The plasma frequency in the ungated structures can be markedly larger
than in the gated GTTSs at the same lengths of the   p- and n-regions. Shortening of these regions is beneficial
for the enhancement of the drag factor $b$ and, hence,  reinforces of the drag related effects. 

Further development of the GTTS-based devices using the Coulomb carrier drag can be realized using periodic structures with
the p$^+$-i-n$^+$ GTTSs connected in series (with  the reverse biased p$^+$-i-n$^+$ GTTSs and forward biased
n$^+$-p$^+$-junctions connecting in series the neighboring GTTSs~\cite{30,31} - the cascade GTTS.                                                                                                                                                                                                                                                                                                                                                                                                                                                                                                                                                                                                                                                                                                                                                                                                                                                                                                                                                                                                                                                                          In such periodic structures one can expect the reinforces THz emission and  the self-excitation of the propagating plasma waves (associated with the plasma instability considered above) with the periods equal to the multiples of the structure period resulting in an additional functionality. 
The vertical integration of the p$^+$-i-n$^+$ GTTSs with the non-Bernal stacked (twisted) G-layers or
G-layers separated by dielectric layers such as hBN layers~\cite{6,32,33} could also promote the enhancement of 
of the GTTS THz radiation sources efficiency (see also~\cite{34}).

\section{Conclusions} 
 
In conclusion,  we predicted  the current driven plasma instability  in
the lateral GTTS with the C injection into the gated p- and n-regions and the Coulomb drag of the ECs by the BCs. The plasma instability and the pertinent emission of the THz radiation  are associated with the current amplification
 due to the transfer of the BC momentum to the ECs.
 The lateral GTTS cascades and e vertically integrated GTTSs  can exhibit an enhanced the THz emission and widen the GTTS functionality. The GTTSs can be effective THz sources surpassing the existing nanostructure emitters.   

The work at RIEC and UoA was supported by the Japan Society for Promotion of Science (KAKENHI Nos.21H04546, 20K20349),
Japan; the RIEC Nation-Wide Collaborative Research
Project No.H31/A01, Japan.  The work at RPI was supported by the Office of Naval Research~(N000141712976,
Project Monitor Dr.~Paul Maki).

\section*{Data Availability}

The data that support the findings of this study are available
from the corresponding author upon reasonable request.

\end{document}